\documentclass[twocolumn,aps,superscriptaddress,showpacs,floatfix]{revtex4}
\usepackage{graphicx}
\usepackage{dcolumn}
\usepackage{bm}
\usepackage[colorlinks,linkcolor=blue,anchorcolor=blue,citecolor=blue,urlcolor=blue]{hyperref}
\usepackage{amsmath}
\usepackage{multirow}
\usepackage{booktabs}  
\usepackage{longtable}  
\usepackage{float}
\usepackage{array}
\usepackage{etoolbox}
\usepackage{makecell,rotating,multirow,diagbox}
\usepackage{setspace}
\usepackage{tabularx}
\usepackage[draft]{changes}
\usepackage[section]{placeins}
\usepackage{lineno,hyperref}
\usepackage{supertabular}
\usepackage{amssymb}
\usepackage[normalem]{ulem}
\usepackage[colorlinks,linkcolor=blue,anchorcolor=blue,citecolor=blue,urlcolor=blue]{hyperref}
\renewcommand\sout{\bgroup \color{red} \ULdepth=-.5ex \ULset}

\begin{document}

\title{Systematic calculations of cluster radioactivity half-lives with a screened electrostatic barrier}

\author{Xiao-Liu}
\affiliation{School of Nuclear Science and Technology, University of South China, 421001 Hengyang, People's Republic of China}
\author{Jie-Dong Jiang}
\affiliation{School of Nuclear Science and Technology, University of South China, 421001 Hengyang, People's Republic of China}
\author{Lin-Jing Qi}
\affiliation{School of Nuclear Science and Technology, University of South China, 421001 Hengyang, People's Republic of China}
\author{Yang-Yang Xu}
\affiliation{School of Nuclear Science and Technology, University of South China, 421001 Hengyang, People's Republic of China}
\author{Xi-Jun Wu}
\email{wuxijunusc@163.com}
\affiliation{School of Math and Physics, University of South China, Hengyang 421001, People's Republic of China}
\author{Xiao-Hua Li}
\email{lixiaohuaphysics@126.com }
\affiliation{School of Nuclear Science and Technology, University of South China, 421001 Hengyang, People's Republic of China}
\affiliation{Cooperative Innovation Center for Nuclear Fuel Cycle Technology $\&$ Equipment, University of South China, 421001 Hengyang, People's Republic of China}
\affiliation{National Exemplary Base for International Sci $\&$ Tech. Collaboration of Nuclear Energy and Nuclear Safety, University of South China, Hengyang 421001, People's Republic of China}

\begin{abstract}
In the present work, based on Wentzel-Kramers-Brillouin theory, we systematically study the cluster radioactivity half-lives of 22 nuclei ranging from $^{221}$$\rm{Fr}$ to $^{242}$$\rm{Cm}$ by using a phenomenological model, which considers the screened electrostatic effect of Coulomb potential. In this model, there are two adjustable parameters i.e. the parameter $t$ and $g$, which are related to the screened electrostatic barrier and the strength of spectroscopic factor, respectively. The calculated results indicate this model can well reproduce the experimental data while the corresponding root-mean-square (rms) deviation is 0.660. In addition, we extend this model to predict the half-lives of possible cluster radioactive candidates whose cluster radioactivity are energetically allowed or observed but not yet quantified in the evaluated nuclear properties table NUBASE2020. The predicted results are consistent with the ones obtained by using other theoretical models and/or empirical formulae including the universal decay law (UDL) proposed by Qi \textit{et al.} [Phys. Rev. C 80, 044326 (2009)], a semi-empirical model for both $\alpha$ decay and cluster radioactivity proposed by Santhosh \textit{et al.} [J. Phys. G 35, 085102 (2008)] and a unified formula of half-lives for $\alpha$ decay and cluster radioactivity proposed by Ni \textit{et al.} [Phys. Rev. C 78, 044310 (2008)]
\end{abstract}

\maketitle

\section{Introduction}
\setlength{\parskip}{0pt}
Spontaneous radioactivity of nucleus has always been one of the front-burner issue in the field of nuclear physics\cite{2000RMP72}. So far, more than 2000 nuclei have been discovered, most of which are unstable\cite{2003NPA7293}, they become stable by spontaneous emission of radiation and/or particles. In the regions of the heavy and superheavy nuclei, cluster radioactivity is one of the main decay modes. In this radioactive process, the unstable parent nucleus emits the cluster particle, which is heavier than the $\alpha$ particle but lighter than the lightest fission fragment. It was firstly predicted  by S$\check{a}$ndulescu, Poenaru and Greiner in 1980\cite{1980WGF111334}. A few years later,  Rose and Jones observed the release of $^{14}$$\rm{C}$ emitted from $^{223}$$\rm{Ra}$ in experiments, which  verified the realistic existence of this radioactivity once again\cite{1984NATURE307245}. From that moment on, cluster radioactivity has evolved as a hot subject and a great deal of works have been devoted to researching and explorating it\cite{1984JPGN10L183,1994PRC491992,1996PRL76380,1998PR294265,2001NPA683182}. Up to now, many types of cluster, such as $^{14}$$\rm{C}$, $^{20}$$\rm{O}$, $^{23}$$\rm{F}$, $^{24,25,26}$$\rm{Ne}$, $^{28,30}$$\rm{Mg}$ and $^{32,34}$$\rm{Si}$ so on, have been discovered and confirmed to be emitted from parent nuclei ranging from $^{221}$$\rm{Fr}$ to $^{242}$$\rm{Cm}$ in experiments\cite{2020PRC102034318,1991JPGN17S443,1985PRC311984,2001NPA68664,1994IJMP3335}. At the same time, during the launch, the parent nuclei decay into the doubly magic daughter nucleus $^{208}$$\rm{Pb}$ or its adjacent daughter nuclei, which aroused the interest of many people because it can provide vital nuclear structure information such as ground state lifetime, nuclear spin and parity, shell effects, nuclear deformations and so on\cite{1989JPGN15529,2007RRP592}.

Theoretically, there are two directions to explain the cluster radioactivity process i.e. spontaneous fission process with super asymmetric mass and $\alpha$-like process.\cite{2003PRC68034321,2008JPGN36015110,2016NPA95186,2020NPA997121714}. In the former, the cluster is considered to be formed continuously with the shape evolution of the parent nucleus as it penetrates the nuclear barrier\cite{2001NPA683182,1991PS44427}. In the latter, as a traditional $\alpha$-decay approach, the cluster is assumed to be preformed in the parent nuclei with a certain probability and then penetrates through the potential barrier\cite{1989PRC391992,1989PRC392097,2009PRC80037307}. 
On the basis of the above theories, extensive theoretical models have been put forward to deal with this type radioactivity such as Coulomb and proximity potential model (CPPM)\cite{2013AP334280}, the preformed cluster model (PCM)\cite{2012PRC86044612,2012PRC85054612}, the generalized liquid drop model (GLDM)\cite{2001NPA683182}, the effective liquid drop model (ELDM)\cite{2022JPGN49025101}, a modified unified fission model (MUFM)\cite{2023CPC47014101}, the density-dependent cluster model (DDCM)\cite{2012JPGN39015103}
and so on\cite{1993PRC482409,2009PRC80024310,2010NPA848292,1989JPGN15615,2005PRC71014301,2013PRC88044608}. The calculated results using these models are basically consistent with the experimental data. At the same time, there are also lots of valid and useful empirical or semi-empirical formulas used to calculate cluster radioactivity half-lives such as the universal decay law (UDL) proposed by Qi \textit{et al.}\cite{2009PRC80044326}, an extension of Viola-Seaborg formula from $\alpha$ decay to cluster radioactivity proposed by Ren \textit{et al.}\cite{2004PRC70034304}, a unified formula of half-lives for $\alpha$ decay and cluster radioactivity proposed by Ni \textit{et al.}\cite{2008PRC78044310},
New Geiger-Nuttall law for $\alpha$ decay and cluster radioactivity half-lives proposed by Ren \textit{et al.}\cite{2012PRC85044608} and so on\cite{2004JPGN30945,2021CPC45044111,2004PRC70017301,2009PRL103072501,2011PRC83014601,2013NPA90840,2018PRC97014318}.

In 2005, Tavares \textit{et al.} firstly proposed a phenomenological one-parameter model(OPM) to evaluated $\alpha$ decay half-lives of bismuth isotopes on account of Wentzel-Kramers-Brillouin theory and considering the overlapping effect\cite{2005JPGN31129}. Since then, the OPM  was successfully applied to evaluate the $\alpha$ decay half-lives of platinum and neptunium isotopes, and the calculated results can reproduce the experimental data well\cite{2006NIMB243256,2021PS96075301}. Recently, Zou \textit{et al.} successfully extended this model in the aspect of proton radioactivity\cite{2022CTP74115302}. Meanwhile, based on the OPM, taking into account the screened electrostatic barrier, Xu \textit{et al.} successfully calculated the $\alpha$ decay half-lives of uranium isotopes\cite{2022CPC46114103}. Since the cluster radioactivity shares the same mechanism of tunneling effect with the proton radioactivity and $\alpha$ decay. Whether the OPM with taking into account the screened electrostatic barrier can be extended to study the cluster radioactivity is an interesting issue. To this end, considering the screened electrostatic effect of Coulomb potential, based on the OPM, we systematically study the cluster radioactivity half-lives of 22 nuclei ranging from $^{221}$$\rm{Fr}$ to $^{242}$$\rm{Cm}$. The calculated results indicate this model can well reproduce the experimental data.

This article is organized as follows. In Section \ref{sec:Theoretical framework}, the theoretical framework for calculating the cluster radioactivity half-life is briefly described. The calculations and discussion are presented in Section \ref{sec:Results and discussion}. Finally, a brief summary is given in Section \ref{sec:Summary}. 

\section{Theoretical framework}
\label{sec:Theoretical framework}
\subsection{The half-lives of the cluster radioactivity}
The cluster radioactivity half-life is defined as\cite{2021IJP95121} 
\begin{eqnarray}\label{1}
{T}_{1/2} = \frac{\rm{ln2}}{\lambda}=\frac{\rm{ln2}}{\nu{S}_{c}{P}},
\end{eqnarray}
where $\lambda$ is the decay constant. It consists of three parts. One of its components, $\nu$ is the assault frequency which represents the number of assaults on the barrier per unit of time. It can be calculated by\cite{2013PRC87024308,2021PS96125322}
\begin{eqnarray}\label{2}
\nu=\frac{\pi\hbar}{2\mu{a}^{2}},
\end{eqnarray}
where $\hbar$ is the reduced Planck constant.
$a=R_{p}-R_{c}$ stands for the inner turning point of potential barrier with $R_{p}$ and $R_{c}$ being the radius of parent nucleus and emitted cluster, respectively. The detailed information about $R_{p}$ and $R_{c}$ will be presented in
next paragraph.
$\mu=\frac{m_{d}m_{c}}{m_{d}+m_{c}}$ denotes the reduced mass between daughter nucleus and emitted cluster. $m_{d}$ and $m_{c}$ are the atomic mass of daughter nucleus and emitted cluster, respectively. In this work, $m_{c}$ is taken from AME2020\cite{2021PCP45030003}, $m_{d}$ can be obtained by
\begin{eqnarray}\label{3}
m_{d}=A_{d}+\frac{\Delta{M}_{d}}{F}-(Z_{d}m_{e}-\frac{10^{-6}kZ_{d}^{\beta}}{F}),
\end{eqnarray}
where $m_{e}$=0.548579911$\times{10^{-3}}u$ is the electron rest mass with $u$ being atomic mass. $F=931.494$MeV $u^{-1}$ is the mass-energy conversion factor. $A_{d}$, $Z_{d}$ and $\Delta{M}_{d}$ are the mass number, proton number and mass excess of the daughter nucleus. The quantity $kZ^{\beta}$ represents the total binding energy of the $Z$ electron in the atom,  the value of $k=8.7$eV, $\beta$=2.517 for $Z\ge60$, while $k=13.6$eV $\beta$=2.408 for $Z<60$\cite{1976ADND18243}.

Another two components, ${S}_{c}$ represents the spectroscopic factor (the probability of finding cluster at the nuclear surface), ${P}$ represents the penetrability factor (the probability of cluster penetrating through external barrier)\cite{2019PPNP105214}, they can be calculated by
\begin{eqnarray}\label{4}
S_{c}=\exp^{-G_{ov}},
G_{ov}=\frac{2}{\hbar}\int_{a}^{b} \sqrt{{2\mu}\left[{V(r)-Q_{c}}\right]}dr,
\end{eqnarray}
\begin{eqnarray}\label{5}
P=\exp^{-G_{se}},
G_{se}=\frac{2}{\hbar}\int_{b}^{c} \sqrt{{2\mu}\left[{V(r)-Q_{c}}\right]}dr.
\end{eqnarray}
Here $G_{ov}$ is the G-factor of the overlapping region, where the cluster particle drives away from the parent nucleus until the outer surface of the nucleus touches each other. $G_{se}$ is the G-factor of the separate region. $b=R_{d}+R_{c}$ and $c$ stand for the separating point and the outer turning point of potential barrier with $R_{d}$ being the radius of daughter nucleus. $R_{p}$ and $R_{d}$ are calculated by the droplet model of atomic nucleus. It can be expressed as\cite{2005JPGN31129,1995ADNDT59185}
\begin{eqnarray}\label{66}
R_{i}=\frac{Z_{i}}{A_{i}}R_{pi}+\left(1-\frac{Z_{i}}{A_{i}}\right) R_{ni}, &i=p,d,
\end{eqnarray}
where $R_{pi}$ and $R_{ni}$ are given by
\begin{eqnarray}\label{67}
R_{ji}=r_{ji}\left[1+\frac{5}{2}\left(\frac{w}{r_{ji}}\right)^{2} \right],&j=p,n,&i=p,d.
\end{eqnarray}
Here $w = 1$ fm is the diffuseness of the nuclear surface.
$r_{ji}$ represent the equivalent sharp radius of
proton $(j = p)$ or neutron $(j = n)$ density distribution of parent nucleus $(i = p)$ or daughter nucleus
$(i = d)$, respectively. According to the finite-range
droplet model theory of nuclei proposed by M$\ddot{o}$ller
$et$ $al$\cite{1995ADNDT59185}, the equivalent sharp radius can be expressed as
\begin{eqnarray}\label{68}
r_{pi}=r_{0}\left(1+\bar{\epsilon_{i}}\right)\left[1-\frac{2}{3}\left(1-\frac{Z_{i}}{A_{i}}\right)\left(1-\frac{2Z_{i}}{A_{i}}-\bar{\delta_{i}}\right)\right]A_{i}^{1/3},
\end{eqnarray}
\begin{eqnarray}\label{69}
r_{ni}=r_{0}\left(1+\bar{\epsilon_{i}}\right)\left[1+\frac{2}{3}\frac{Z_{i}}{A_{i}}\left(1-\frac{2Z_{i}}{A_{i}}-\bar{\delta_{i}}\right)\right]A_{i}^{1/3},
\end{eqnarray}
where $r_{0}=1.16$ fm, $i = p$ (parent nucleus) or $d$ (daughter nucleus). $\bar{\epsilon_{i}}$ and $\bar{\delta_{i}}$ can be given by
\begin{eqnarray}\label{70}
\bar{\epsilon_{i}}=\frac{1}{4e^{0.831A_{i}^{1/3}}}-\dfrac{0.191}{A_{i}^{1/3}}+\frac{0.0031Z_{i}^{2}}{A_{i}^{4/3}},
\end{eqnarray}
\begin{eqnarray}\label{71}
\bar{\delta_{i}}=\left(1-\frac{2Z_{i}}{A_{i}}+0.004781\frac{Z_{i}}{A_{i}^{2/3}}\right)\mathbf{/}\left(1+\frac{2.52114}{A_{i}^{1/3}}\right).
\end{eqnarray}
The above radius parametrization does not apply to lighter nuclei with $Z<8$ and $A<16$\cite{1995ADNDT59185,2018EPJA5465}. For unity, the radius of all emitted clusters is parameterized as following\cite{2022CTP74115302}.
\begin{eqnarray}\label{72}
R_{c}=1.28A_{c}^{1/3}-0.76+0.8A_{c}^{-1/3},
\end{eqnarray}
where $A_{c}$ is the mass number of emitted cluster.

\begin{figure}[h]\centering
	\includegraphics[width=9cm]{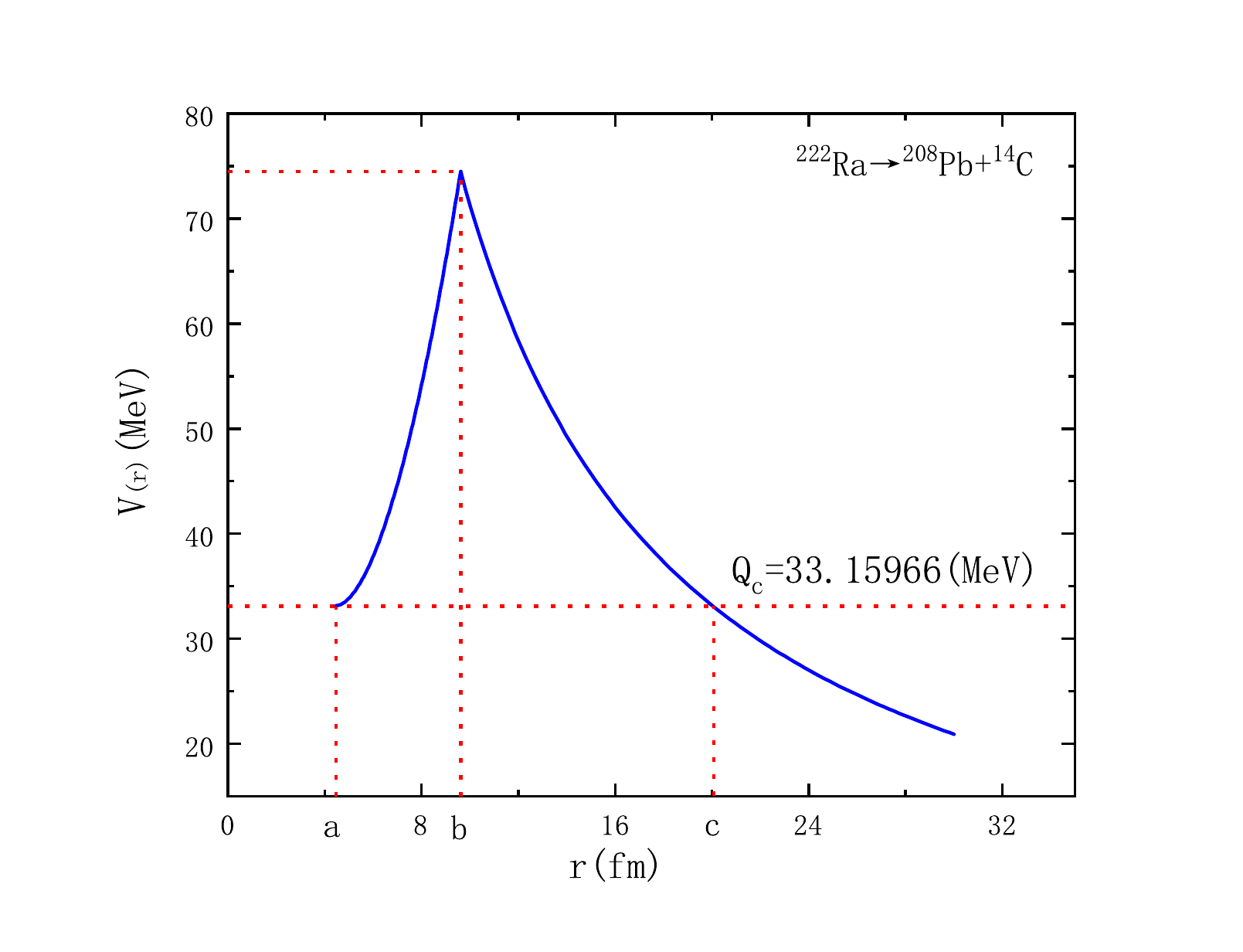}
	\caption{Schematic diagram of interaction
		potential $V(r)$ between cluster and daughter nucleus. $a$, $b$ and $c$ are the three turning points of the potential energy while the barrier of $a$ to $b$ is overlapping region and $b$ to $c$ is the separation region.}
	\label{fig1}
\end{figure}

$Q_{c}$ represents the cluster radioactivity released energy. It can be obtained by
\begin{eqnarray}\label{6}
Q_{c}=\Delta{M}_{p}-(\Delta{M}_{d}+\Delta{M}_{c})+10^{-6}k(Z_{p}^{\beta}-Z_{d}^{\beta}),
\end{eqnarray}
where $\Delta{M}_{p}$ and $\Delta{M}_{c}$ are the mass excess of parent nucleus and emitted cluster, respectively. They are extracted from the evaluated nuclear properties table NUBASE2020\cite{2021CPC45030001}. The term $k(Z_{p}^{\beta}-Z_{d}^{\beta})$ represents the screened effect of the atomic electrons, the value of $k=8.7$eV, $\beta$=2.517 for $Z\ge60$, while $k=13.6$eV $\beta$=2.408 for $Z<60$\cite{1976ADND18243}.  
$V(r)$ roughly shown in Fig. \ref{fig1} is the total interaction potential between daughter nucleus and emitted cluster. At two points $a$ and $c$, they satisfy the condition $V(a)=V(c)=Q_{c}$.
In general, the total interaction potential $V(r)$ consists of 
the nuclear potential $V_{N}(r)$, Coulomb potential $V_{C}(r)$
and centrifugal potential $V_{\ell}(r)$. It can be expressed as
\begin{eqnarray}\label{7}
V(r)=V_{N}(r)+V_{C}(r)+V_{\ell}(r).
\end{eqnarray}

In this work, for the centrifugal potential $V_{\ell}(r)$, we adopt the Langer modified centrifugal potential
\cite{2020JPGN47055105}, because ${\ell}({\ell+1})\to({\ell+\dfrac{1}{2}})^{2}$ is a necessary correction for one-dimensional problems. It can be expressed as
\begin{eqnarray}\label{8}
V_{\ell}(r)=\frac{\left(\ell+1/2\right)^{2}\hbar^{2}}{2\mu{r}^{2}},
\end{eqnarray}
where $\ell$ is the angular momentum carried by the emitted cluster. It can be obtained by
\begin{eqnarray}\label{9}
\ell=\left\{\begin{array}{llll}
\Delta_{j}, &$for even $\Delta_{j}$ and $\pi_{p} = \pi_{d}$$,\\
\Delta_{j}+1, &$for even $\Delta_{j}$ and $\pi_{p} \neq \pi_{d}$$,\\
\Delta_{j}, &$for odd $\Delta_{j}$ and $\pi_{p} \neq \pi_{d}$$,\\
\Delta_{j}+1, &$for odd $\Delta_{j}$ and $\pi_{p} = \pi_{d}$$,
\end{array}\right.
\end{eqnarray}
where $\Delta_{j}$ = $\lvert {j_{p}-j_{d}-j_{c}} \rvert $, $j_{p}$, $\pi_{p}$, $j_{d}$, $\pi_{d}$ and $j_{c}$, $\pi_{c}$ represents spin and parity values of parent, daughter and cluster nuclei, respectively. They are taken from NUBASE2020\cite{2021CPC45030001}.

For the Coulomb potential $V_{C}(r)$, it is usually defined as  
\begin{eqnarray}\label{10}
V_{C}(r)=\frac{Z_{c}Z_{d}e^{2}}{r},
\end{eqnarray}
where $e^{2}=1.4399652$ MeV$\cdot$fm is the square of the electronic elementary charge. Since the process of cluster radioactivity process involves the superposition of the involved charges, the inhomogeneous charge distribution of the nucleus and movement of the emitted cluster which generates a magnetic field. Therefore the electrostatic potential of the emitted cluster-daughter nucleus behaves as a Coulomb potential at short distances and drops exponentially at large distance i.e. the screened electrostatic effect. This behaviour of electrostatic potential can be described as the Hulthen type potential $V_{H}(r)$. It is defined as\cite{1942AMAFA2852,2020CPC124102,2019NPA987350}
\begin{eqnarray}\label{11}
V_{H}(r)=\frac{tZ_{c}Z_{d}e^{2}}{e^{tr}-1},
\end{eqnarray}
where $t$ is the screening parameter.

In the overlapping region, $\mu$ and $V(r)$ can not be calculated as a oversimplified two-body system since the cluster is still in the parent nucleus. In order to deal with this problem accurately, according to Ref.\cite{2005JPGN31129,1996MGGP532309,1985PRC32572}, $\mu$ and $V(r)$ are rewritten as the following form
\begin{eqnarray}\label{12}
\mu(r)=\left( \frac{m_{d}m_{c}}{m_{d}+m_{c}}\right) \left( \frac{r-a}{b-a}\right)^{p},p\ge{0},
\end{eqnarray}
\begin{eqnarray}\label{13}
&V(r)=Q_{c}+\left( V(b)-Q_{c}\right) \left( \frac{r-a}{b-a}\right)^{q},q\ge{1},
\end{eqnarray}
with
\begin{eqnarray}\label{14}
V(b)=\frac{tZ_{c}Z_{d}e^{2}}{e^{tb}-1}+\frac{\left(\ell+1/2\right)^{2}\hbar^{2}}{2\mu{b}^{2}}.
\end{eqnarray}
Using the Eq.(\ref{4}) (\ref{12}) (\ref{13}) (\ref{14}), the $G_{ov}$ can be expressed as 
\begin{eqnarray}\label{15}
&G_{ov}=\sqrt{\frac{931.494}{9\times10^{46}}}\frac{2\sqrt{2}(b-a)g}{\hbar}\times\nonumber\\
&\sqrt{\mu\left[\frac{tZ_{c}Z_{d}e^{2}}{e^{tb}-1}+\frac{9\times10^{46}}{931.494}\frac{\hbar^{2}(\ell+1/2)^{2}}{2\mu{b}^{2}}-Q_{c}\right]}\nonumber\\
&=0.4374702(b-a)g\times\nonumber\\
&\sqrt{\mu\left[\frac{tZ_{c}Z_{d}e^{2}}{e^{tb}-1}+\frac{20.9008(\ell+1/2)^{2}}{\mu{b}^{2}}-Q_{c}\right]},
\end{eqnarray}
where $g=(1+\frac{p+q}{2})^{-1}$ is related to the strength of the spectroscopic factor with $0\le{g}\le{\frac{2}{3}}$. In the separated region, cluster is separated from the parent nucleus and the whole system is regarded as a simple two-boby system. According to $\mu=\frac{m_{d}m_{c}}{m_{d}+m_{c}}$ and Eq.(\ref{5})(\ref{7}) (\ref{8}) (\ref{11}), the $G_{se}$ can be obtained as
\begin{eqnarray}\label{16}
\begin{aligned}
G_{se}&=\sqrt{\frac{931.494}{9\times10^{46}}}\frac{2\sqrt{2}e^{2}{Z_{c}}{Z_{d}}}{\hbar}\sqrt{\dfrac{\mu}{Q_{c}}}\times{F}\\
&=0.62994397{Z_{c}}{Z_{d}}\sqrt{\dfrac{\mu}{Q_{c}}}\times{F},
\end{aligned}
\end{eqnarray}
where
\begin{eqnarray}\label{17}
&F=\frac{x^{1/2}}{2y}\times\ln\left(\dfrac{\sqrt{x(x+2y-1)}+x+y}{(x/y)\left[1+\sqrt{1+x/y^{2}}\right]^{-1}+y}\right)\nonumber\\
&+\arccos\sqrt{\frac{1}{2}\left(1-\frac{1-1/y}{\sqrt{1+x/y^2}}\right)}\nonumber\\
&-\sqrt{\dfrac{1}{2y}(1+x/2y-1/2y)},
\end{eqnarray}
with
\begin{eqnarray}\label{18}
x=\frac{9\times10^{46}}{931.494}\frac{\hbar^{2}(\ell+1/2)^{2}}{2\mu{b}^{2}Q_{C}}=\frac{20.9008(\ell+1/2)^{2}}{\mu{b}^{2}Q_{C}},
\end{eqnarray}
\begin{eqnarray}\label{73}
y=\dfrac{\ln(tZ_{c}Z_{d}e^{2}/Q_{c}+1)}{2tb}.
\end{eqnarray}

Based on the above, the cluster radioactivity half-life can be expressed as
\begin{eqnarray}\label{19}
\begin{aligned}
T_{1/2}&=\frac{931.494}{9\times10^{46}}\times\frac{2\ln{2}}{\pi\hbar}\mu{a}^{2}S_{c}^{-1}P^{-1}\\
&=6.93868752\times10^{-24}\mu{a}^{2}S_{c}^{-1}P^{-1}.
\end{aligned}
\end{eqnarray}

\subsection{Empirical and semi-empirical formula}

\subsubsection{Universal decay law}

In 2009, based on $\alpha$-like $R$-matrix theory, Qi \textit{et al.} proposed the universal decay law (UDL)  \cite{2009PRC80044326,2013EPJA4966}. It can be expressed as
\begin{eqnarray}\label{20}
\rm{log}_{10}{\textit{T}_{1/2}}=\textit{a}Z_{c}Z_{d}\sqrt{\frac{\mathcal{A}}{Q_{c}}}+b\sqrt{\mathcal{A}Z_{c}Z_{d}(A_{c}^{1/3}+A_{d}^{1/3})}+c,
\end{eqnarray}
where $\mathcal{A}=A_{c}A_{d}/(A_{c}+A_{d})$ is the reduced mass of the emitted cluster-daughter nucleus system measured in unit of the
nucleon mass with $Z_{d}$, $A_{d}$ and $Z_{c}$,  $A_{c}$ being the  proton and mass number of daughter and cluster nucleus, respectively. $Q_{c}$ represents the cluster radioactivity released energy. $a=0.4314$, $b=-0.3921$ and $c=-32.7044$ are the adjustable parameters.

\subsubsection{Santhosh's semi-empirical formula}

In 2008, based on the Geiger-Nuttall(G-N) law, Santhosh \textit{et al.} proposed a semi-empirical model to calculate the half-lives of $\alpha$ decay and cluster radioactivity\cite{2008JPGN35085102}. It can be expressed as

\begin{eqnarray}\label{21}
\rm{log}_{10}{\textit{T}_{1/2}}=\textit{a}Z_{c}Z_{d}Q_{c}^{-1/2}+b\eta_{A}+c,
\end{eqnarray}
where $\eta_{A}=\frac{A_{d}-A_{c}}{A}$ is the mass asymmetry. $Q_{c}$ represents $\alpha$ decay energy and cluster radioactivity released energy. $a=0.727356$, $b=40.3887$ and $c=-85.1625$ are the adjustable parameters.

\subsubsection{Ni's empirical formula}

In 2008, Ni \textit{et al.} proposed a unified formula of half-lifes for $\alpha$ decay and cluster radioactivity by WKB approximation\cite{2008PRC78044310}. It can be expressed as

\begin{eqnarray}\label{22}
\rm{log}_{10}{\textit{T}_{1/2}}=\textit{a}\sqrt{\mathcal{A}}Z_{c}Z_{d}Q_{c}^{-1/2}+b\sqrt{\mathcal{A}}(Z_{c}Z_{d})^{1/2}+c,
\end{eqnarray}
where $\mathcal{A}=A_{c}A_{d}/(A_{c}+A_{d})$ is the reduced mass the same as UDL. $Q_{c}$ represents $\alpha$ decay energy and cluster radioactivity released energy. $a=0.38617$, $b=-1.08676$, $c_{e-e}=-21.37195$ and $c_{o-e,o-e}=-20.11223$ are the adjustable parameters.

\section{Results and discussion}
\label{sec:Results and discussion}
Based on Wentzel-Kramers-Brillouin theory, considering the screened electrostatic effect of Coulomb potential, we propose a phenomenological model by modifying the OPM to calculate the cluster radioactivity half-lives of 22 nuclei, which contains two adjustable parameters i.e. the parameter $t$ and $g$. The standard deviation between the experimental data and calculated ones is taken as the objective function by using the 22 nuclei experimental data in the genetic algorithm, we obtain the optimal adjustable parameters $g=0.416617$, $t=9.176915\times10^{-3}$. In this work, the standard deviation $\sigma$ is defined as
\begin{eqnarray}\label{23}
\sigma = \sqrt{\sum{(\rm{log}_{10}{\textit{T}_{1/2}^{\rm{cal}}}-\rm{log}_{10}{\textit{T}_{1/2}^{\rm{exp}}})^2}/n},
\end{eqnarray}
where $\rm{log}_{10}{\textit{T}_{1/2}^{\rm{exp}}}$ and $\rm{log}_{10}{\textit{T}_{1/2}^{\rm{cal}}}$ are logarithmic form of experimental cluster radioactivity half-lives and the calculated ones, $n$ is the number of nuclei involved for different decay cases. We calculate the standard deviation $\sigma$ between the experimental data and calculated ones obtained by using our model, OPM, UDL, Santhosh's semi-empirical formula and Ni's formula denoted as $\sigma_{Cal}$, $\sigma_{OPM}$, $\sigma_{UDL}$, $\sigma_{San}$ and $\sigma_{Ni}$, respectively. The results are shown in Table \ref{Tab1}. From this Table, we can clearly see the standard deviation $\sigma$ of our result is 0.660. It is better than OPM, UDL and Santhosh's semi-empirical formula results of 0.904, 1.237 and 2.470, respectively, but worse than the Ni's formula result of 0.456. In particular, the $\sigma$ value for 22 nuclei within modified OPM drops $\frac{0.904-0.660}{0.904}=26.99\%$ relative to the one within OPM. It indicates that the calculated half-lives given by our modified OPM can better reproduce the experimental data.

\begin{table}[!htb] 
	\renewcommand\arraystretch{1.5}
	\caption{Standard deviation $\sigma$ between experimental half-lives and the calculated ones obtained by using different formulas and/or models for cluster radioactivity.} 
	\label{Tab1} 
	\centering
	\begin{tabular*}{8cm} {@{\extracolsep{\fill}} |c|c|c|c|c|c|}
		\hline
		\diagbox{cases}{model}&$\sigma_{Cal}$&$\sigma_{OPM}$&$\sigma_{UDL}$&$\sigma_{San}$&$\sigma_{Ni}$ 
		\\  
		\hline 
		22 nuclei  &0.660 &0.904 &1.237 &2.470 & 0.456
	    \\	 
		\hline
	\end{tabular*}  
\end{table}

\begin{figure}[h]\centering
	\includegraphics[width=9cm]{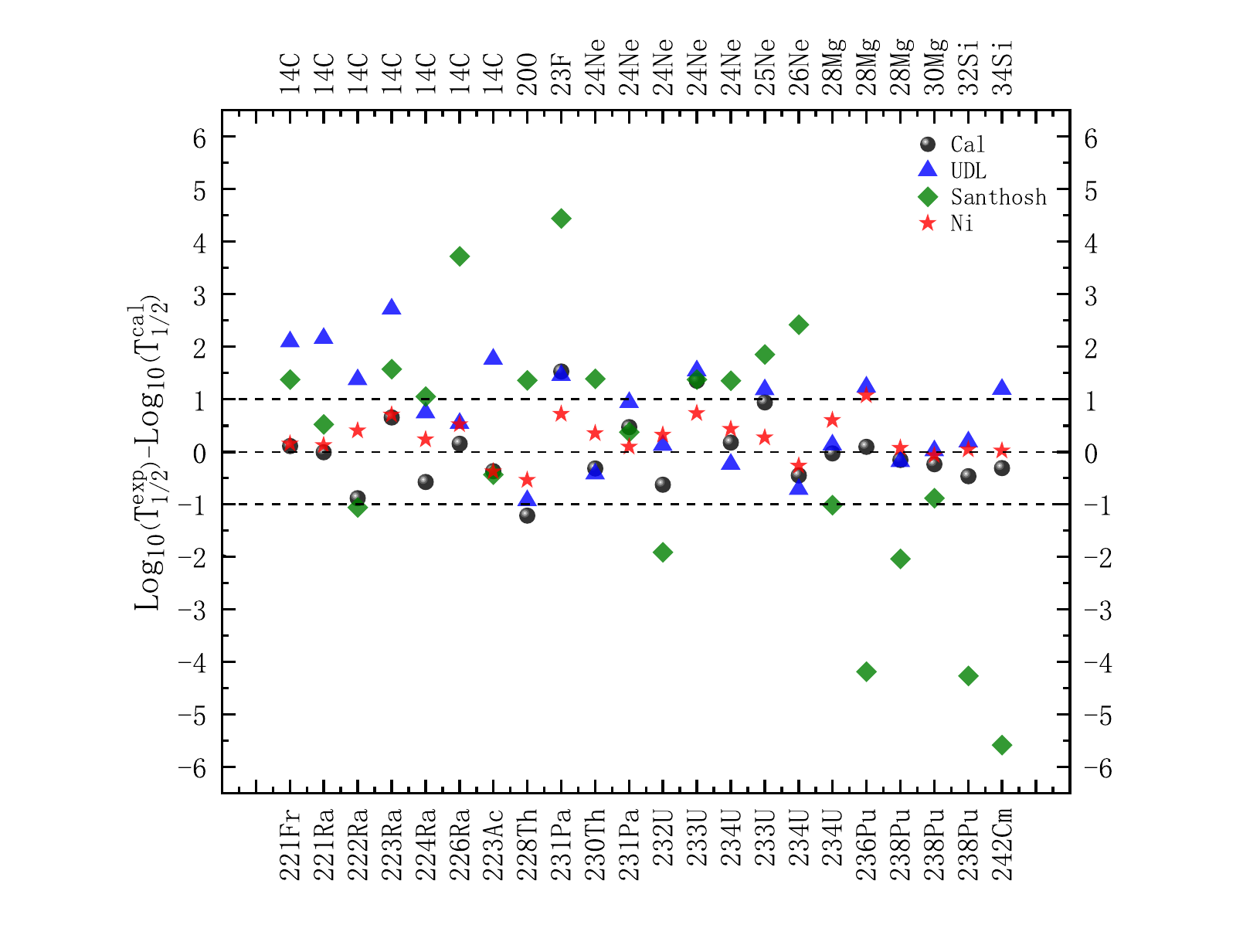}
	\caption{(color online) Comparison of the discrepancy between the experimental cluster radioactivity half-lives and calculated ones obtained in our model, UDL, Santhosh's semi-empirical formula and Ni's formula in logarithmical form.}
	\label{fig2}
\end{figure}

Furthermore, using the optimal parameter values of $t$  and $g$, based on our modified OPM, we systematically calculate the cluster radioactivity half-lives of 22 nuclei. For comparision, UDL, Santhosh's semi-empirical formula and Ni's formula are also used.  All detailed results are given in Table \ref{Tab2}. In this table, the first to third columns are the decay process of the cluster, the angular momentum $l$ taken away by the emitted cluster, the cluster radioactivity released energy $Q_{c}$, respectively. The fourth and fifth columns are the spectroscopic factor $S_{c}$ and penetrability factor $P$ in logarithmical form. The sixth to tenth columns are experimental data of cluster radioactive half-lifes and the calculated ones by using our model, UDL, Santhosh's semi-empirical formula and Ni's formula in logarithmic form denoted as 
$\rm{log}_{10}{\textit{T}_{1/2}^{\rm{exp}}}$, $\rm{log}_{10}{\textit{T}_{1/2}^{\rm{cal}}}$, $\rm{log}_{10}{\textit{T}_{1/2}^{\rm{UDL}}}$, $\rm{log}_{10}{\textit{T}_{1/2}^{\rm{San}}}$ and $\rm{log}_{10}{\textit{T}_{1/2}^{\rm{Ni}}}$, respectively.
As we can see from this table, the calculated half-lives using our model are in good agreement with experimental data as well as the calculated results given by using Ni's formula. To intuitively comparing these results, we plot the differences between the experimental cluster radioactivity half-lives and the calculated ones by using our model, UDL, Santhosh's semi-empirical formula and Ni's formula in logarithmical form in Fig. \ref{fig2}.
\begin{figure}[h]\centering
	\includegraphics[width=9cm]{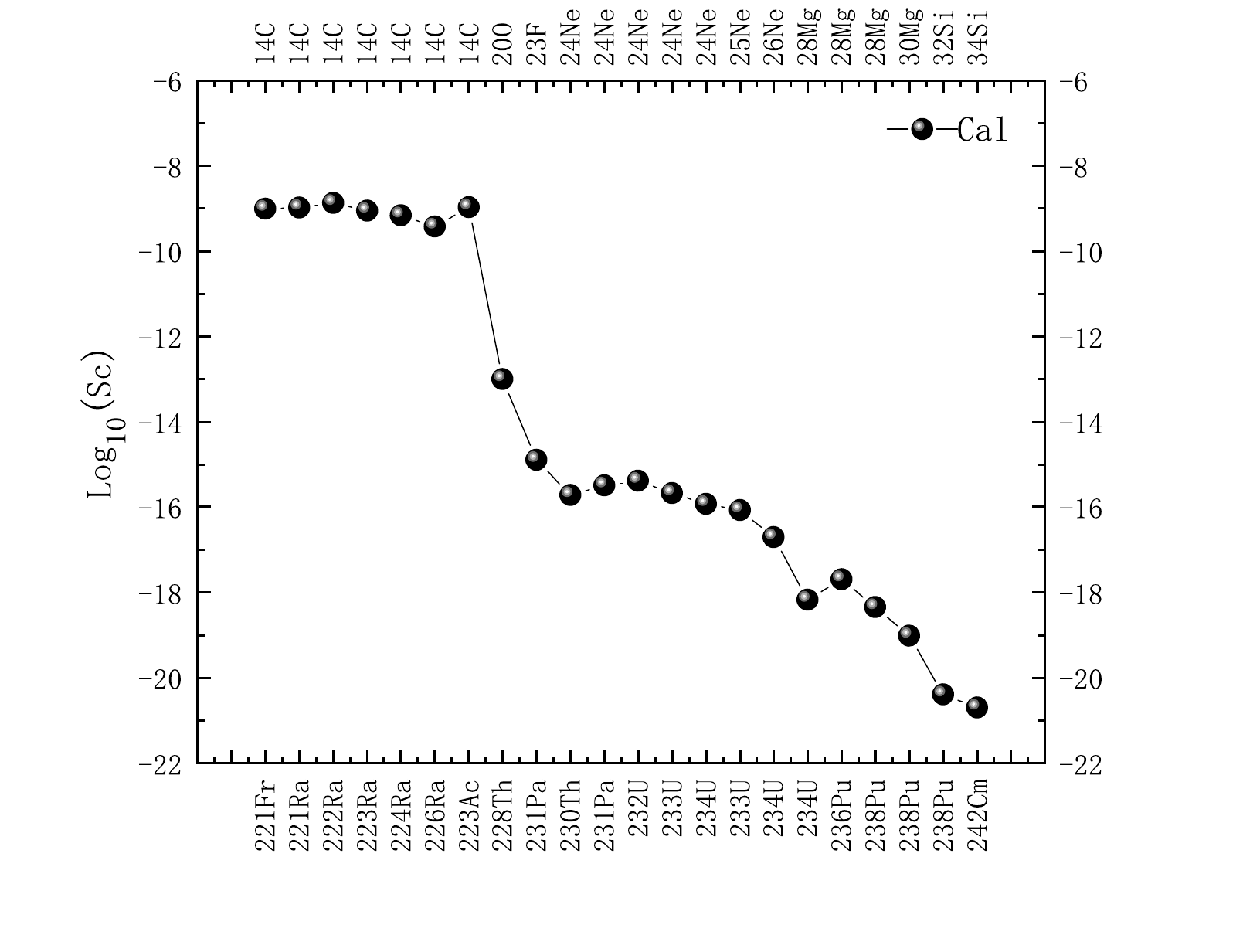}
	\caption{The relationship between the preformation probability calculated by our model in logarithmical form and corresponding 22 cluster radioactivity nuclei.}
	\label{fig3}
\end{figure}
From this figure, we can clearly see that the differences between the experimental data and the calculated results by our model and Ni's formula are within or near $\pm1$ on the whole. For UDL, there is three nuclei beyond the scale of $\pm2$ while it is up to seven nuclei for Santhosh's semi-empirical formula. In its case of $^{231}\rm{Pa}\to^{208}\rm{Pb} +^{23}\rm{F}$ and $^{242}\rm{Cm}\to^{208}\rm{Pb}+^{34}\rm{Si}$, the discrepancies can even achieve at 4.44 and 5.59, respectively. Furthermore, from the overall trend, the results in our model and Ni's formula are more converging on the neighboring zero line area while the distribution for the results of Santhosh's semi-empirical formula is slightly scattered.
Based on the above, we can find that the cluster radioactivity half-life calculated by our model can well reproduce the experimental data. 
\begin{figure}[h]\centering
	\includegraphics[width=9cm]{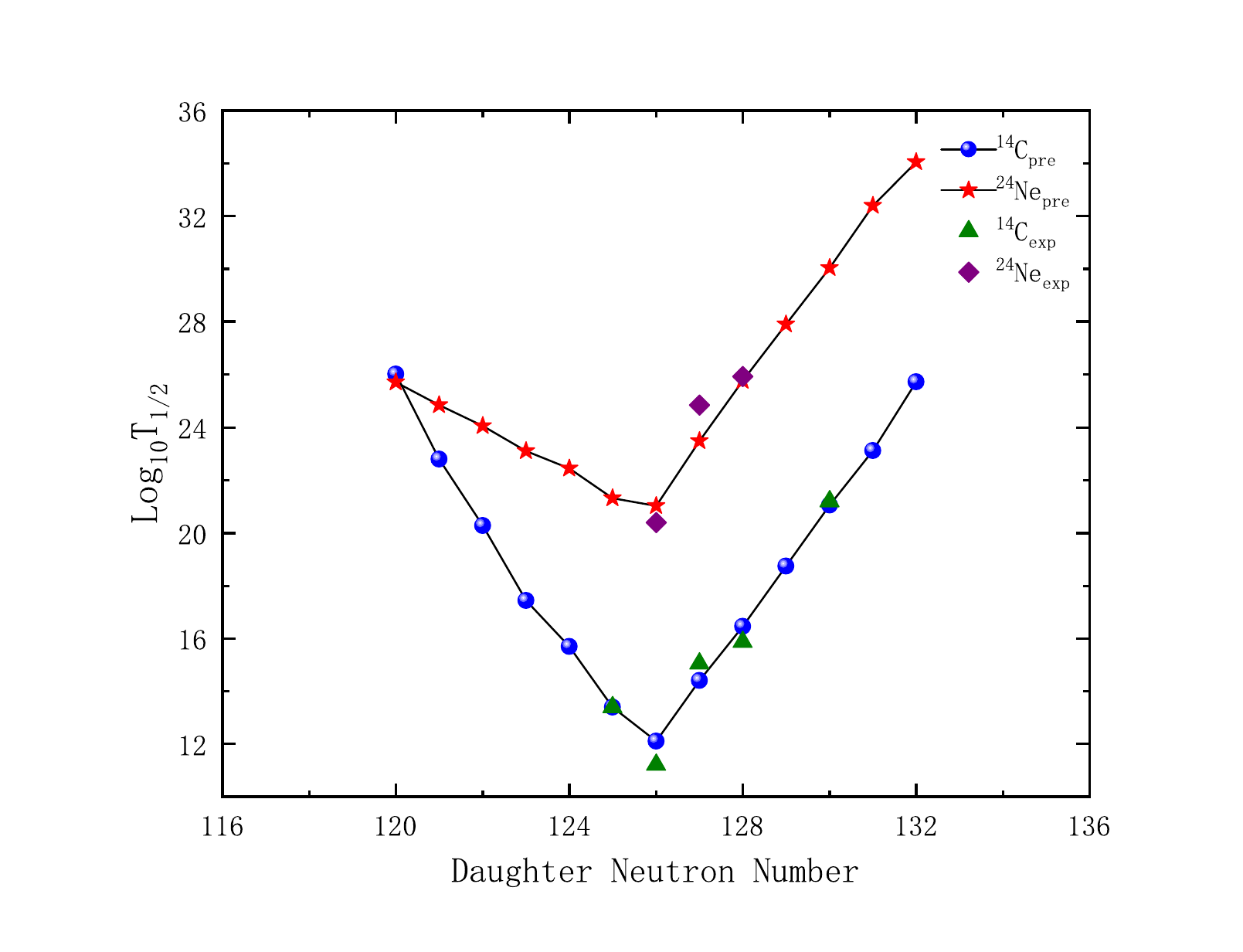}
	\caption{(color online) Plot of the calculated $\rm{log}_{10}{\emph{T}}_{1/2}$ values by our model versus neutron number of daughter for the emission of cluster $ ^{14}\rm{C}$ from $ \rm{Ra} $ isotopes and the emission of cluster $ ^{24}\rm{Ne}$ from $ \rm{U} $ isotopes.}
	\label{fig4}
\end{figure}
\begin{figure}[h]\centering
	\includegraphics[width=9cm]{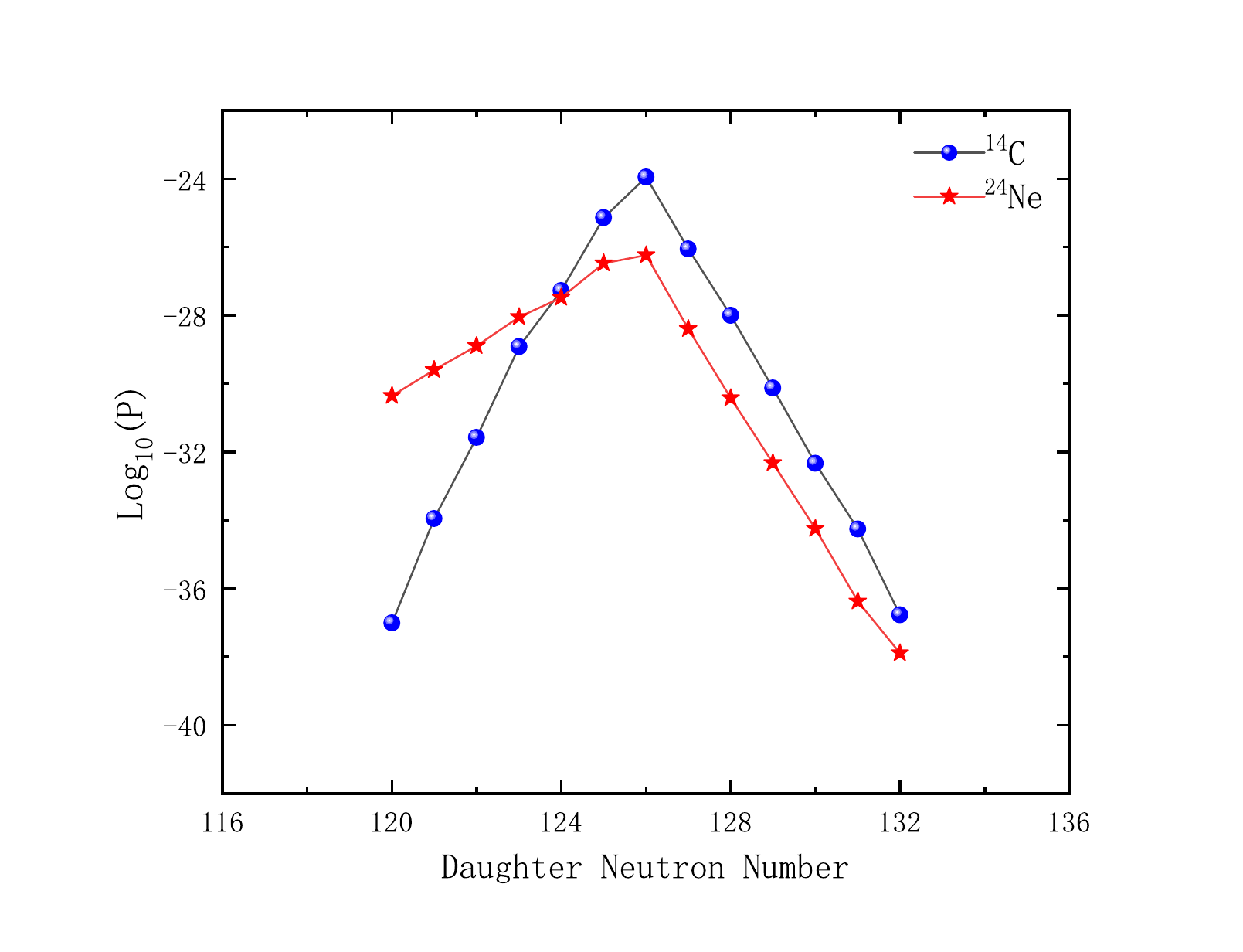}
	\caption{(color online) Plot of the calculated $\rm{log}_{10}{\emph{P}}$ values by our model versus neutron number of daughter for the emission of cluster $ ^{14}\rm{C}$ from $ \rm{Ra} $ isotopes and the emission of cluster $ ^{24}\rm{Ne}$ from $ \rm{U} $ isotopes.}
	\label{fig5}
\end{figure}

In addition, we draw the relationship between 22 cluster radioactivity nuclei and the corresponding preformation probability calculated by our model in logarithmic form in Fig. \ref{fig3}. In this figure, the preformation probability value is found to decrease as the size of the cluster increases. For different cluster radioactivity nuclei, most of the preforming probability of emitting clusters with the same number of protons is consistent. The structure of this curve matches well with the other preformation probability values from Ref.\cite{2018EPJA541}. Nevertheless the results of their calculations vary greatly, which may be the model dependent reason. The nucleon pairing effect also has an important impact on the radioactivity of nuclei. In order to understand it in depth, we plot the relationship between the cluster radioactivity half-lifes, penetration probability in logarithmic form and neutron number of corresponding daughter nuclei in Fig. \ref{fig4} and Fig. \ref{fig5}, respectively. From Fig. \ref{fig4}, it is obvious that for different emitted clusters, $\rm{log}_{10}{\emph{T}}_{1/2}$ first linearly decrease with the increase of neutron number of daughter nuclei, then reach the minimum value when it drops to the magic number of neutron $(N_{d}=126)$, final linearly increase. The tendency is exactly opposite for the $\rm{log}_{10}{\emph{P}}$ depicted in Fig. \ref{fig5}. However, they both have a turning trend when daughter nucleus reaches the double magic number nucleus $ ^{208}\rm{Pb}$, at which point they obtain the minimum and maximum values. Consequently, it indicates the neutron shell closure plays an more important role than the proton shell closure in the process of cluster radioactivity\cite{2020PS95075303}.

Encouraged by the good agreement between the experimental cluster radioactivity half-lives and calculated ones by using our model,  we extend this model to predict the half-lives for the possible cluster radioactive candidates. The detailed predictions are shown in the Table \ref{Tab3}. In this table, the meaning of each column is exactly the same as table \ref{Tab2}. From the Table \ref{Tab3}, it is obviously to see that predicted results of our model are basically consistent with the ones obtained by using UDL and Ni's formula. This shows that our model is reliable for calculating cluster radioactivity. Meanwhile, these predicted values will be helpful for searching new cluster emitters in future experiments.

\begingroup
\renewcommand*{\arraystretch}{1.3}
\setlength{\tabcolsep}{1mm}
\setlength\LTleft{0pt}
\setlength\LTright{0pt}
\setlength{\LTcapwidth}{7in}
\begin{longtable*}
	{@{\extracolsep{\fill}} cccccccccc}
	\caption{Comparison of the discrepancy in logarithmical form between the experimental cluster radioactivity half-lives and calculated ones by using our model, UDL, Santhosh's semi-empirical formula and Ni's formula. The cluster released energy and half-lives are in the unit of $(MeV)$ and $(s)$, respectively. The experimental cluster radioactivity half-lives are taken from Ref.\cite{2023CPC47014101,2010PRC024311,2015MPLA301550150}.}
	\label{Tab2} \\
	\hline 
	\hline 
	{Cluster decay}&$\ell$&$Q_{c}$&$S_{c}$&$P$&
	$\rm{log}_{10}{\emph{T}}_{1/2}^{\rm{\,exp}}$&$\rm{log}_{10}{\emph{T}}_{1/2}^{\rm{\,cal}}$&$\rm{log}_{10}{\emph{T}}_{1/2}^{\rm{\,UDL}}$&$\rm{log}_{10}{\emph{T}}_{1/2}^{\rm{\,San}}$&$\rm{log}_{10}{\emph{T}}_{1/2}^{\rm{\,Ni}}$
	\\
	\hline
	\endfirsthead
	\multicolumn{10}{c}%
	{{\tablename\ \thetable{} continued from previous page}} \\
	\hline
	\hline 
    {Cluster decay}&$\ell$&$Q_{c}$&$S_{c}$&$P$&
    $\rm{log}_{10}{\emph{T}}_{1/2}^{\rm{\,exp}}$&$\rm{log}_{10}{\emph{T}}_{1/2}^{\rm{\,cal}}$&$\rm{log}_{10}{\emph{T}}_{1/2}^{\rm{\,UDL}}$&$\rm{log}_{10}{\emph{T}}_{1/2}^{\rm{\,San}}$&$\rm{log}_{10}{\emph{T}}_{1/2}^{\rm{\,Ni}}$	\\
	\hline
	\endhead
	\hline  \\
	\endfoot
	\hline \hline
	\endlastfoot	
	$^{221}$Fr$\to^{207}$Tl$+^{14}$C$$&$3$&$	31.4002 	$&$	-9.0144 	$&$	-26.1482 	$&$	14.56 	$&$	14.45 	$&$	12.47 	$&$	13.19 	$&$	14.41 	$\\
	$^{221}$Ra$\to^{207}$Pb$+^{14}$C$$&$3$&$	32.5072 	$&$	-8.9798 	$&$	-25.1296 	$&$	13.39 	$&$	13.40 	$&$	11.23 	$&$	12.87 	$&$	13.27 	$\\
	$^{222}$Ra$\to^{208}$Pb$+^{14}$C$$&$0$&$	33.1597 	$&$	-8.8685 	$&$	-23.9449 	$&$	11.22 	$&$	12.11 	$&$	9.85 	$&$	12.28 	$&$	10.82 	$\\
	$^{223}$Ra$\to^{209}$Pb$+^{14}$C$$&$4$&$	31.9390 	$&$	-9.0515 	$&$	-26.0547 	$&$	15.05 	$&$	14.40 	$&$	12.33 	$&$	13.48 	$&$	14.35 	$\\
	$^{224}$Ra$\to^{210}$Pb$+^{14}$C$$&$0$&$	30.6454 	$&$	-9.1609 	$&$	-27.9956 	$&$	15.87 	$&$	16.45 	$&$	15.13 	$&$	14.82 	$&$	15.64 	$\\
	$^{226}$Ra$\to^{212}$Pb$+^{14}$C$$&$0$&$	28.3077 	$&$	-9.4243 	$&$	-32.3293 	$&$	21.20 	$&$	21.05 	$&$	20.66 	$&$	17.48 	$&$	20.68 	$\\
	$^{223}$Ac$\to^{209}$Bi$+^{14}$C$$&$2$&$	33.1768 	$&$	-8.9726 	$&$	-24.6986 	$&$	12.60 	$&$	12.97 	$&$	10.84 	$&$	13.04 	$&$	12.98 	$\\
	$^{228}$Th$\to^{208}$Pb$+^{20}$O$$&$0$&$	44.8739 	$&$	-13.0015 	$&$	-29.5698 	$&$	20.73 	$&$	21.95 	$&$	21.66 	$&$	19.37 	$&$	21.27 	$\\
	$^{231}$Pa$\to^{208}$Pb$+^{23}$F$$&$1$&$	52.0539 	$&$	-14.8949 	$&$	-30.1885 	$&$	26.02 	$&$	24.49 	$&$	24.57 	$&$	21.58 	$&$	25.30 	$\\
	$^{230}$Th$\to^{206}$Hg$+^{24}$Ne$$&$0$&$	57.9453 	$&$	-15.7237 	$&$	-29.8159 	$&$	24.63 	$&$	24.95 	$&$	25.05 	$&$	23.24 	$&$	24.28 	$\\
	$^{231}$Pa$\to^{207}$Tl$+^{24}$Ne$$&$1$&$	60.5984 	$&$	-15.4928 	$&$	-27.5145 	$&$	22.89 	$&$	22.42 	$&$	21.95 	$&$	22.52 	$&$	22.80 	$\\
	$^{232}$U$\to^{208}$Pb$+^{24}$Ne$$&$0$&$	62.5013 	$&$	-15.3769 	$&$	-26.2290 	$&$	20.39 	$&$	21.02 	$&$	20.27 	$&$	22.31 	$&$	20.07 	$\\
	$^{233}$U$\to^{209}$Pb$+^{24}$Ne$$&$2$&$	60.6771 	$&$	-15.6747 	$&$	-28.3978 	$&$	24.84 	$&$	23.49 	$&$	23.30 	$&$	23.47 	$&$	24.11 	$\\
	$^{234}$U$\to^{210}$Pb$+^{24}$Ne$$&$0$&$	59.0168 	$&$	-15.9235 	$&$	-30.4173 	$&$	25.93 	$&$	25.76 	$&$	26.17 	$&$	24.58 	$&$	25.50 	$\\
	$^{233}$U$\to^{208}$Pb$+^{25}$Ne$$&$2$&$	60.8954 	$&$	-16.0720 	$&$	-28.4021 	$&$	24.84 	$&$	23.90 	$&$	23.66 	$&$	22.99 	$&$	24.57 	$\\
	$^{234}$U$\to^{208}$Pb$+^{26}$Ne$$&$0$&$	59.6043 	$&$	-16.7010 	$&$	-30.2412 	$&$	25.93 	$&$	26.38 	$&$	26.65 	$&$	23.51 	$&$	26.20 	$\\
	$^{234}$U$\to^{206}$Hg$+^{28}$Mg$$&$0$&$	74.3372 	$&$	-18.1723 	$&$	-27.9477 	$&$	25.53 	$&$	25.56 	$&$	25.40 	$&$	26.55 	$&$	24.93 	$\\
	$^{236}$Pu$\to^{208}$Pb$+^{28}$Mg$$&$0$&$	79.9042 	$&$	-17.6878 	$&$	-24.2891 	$&$	21.52 	$&$	21.43 	$&$	20.29 	$&$	25.71 	$&$	20.45 	$\\
	$^{238}$Pu$\to^{210}$Pb$+^{28}$Mg$$&$0$&$	76.1457 	$&$	-18.3360 	$&$	-28.0659 	$&$	25.70 	$&$	25.86 	$&$	25.89 	$&$	27.74 	$&$	25.63 	$\\
	$^{238}$Pu$\to^{208}$Pb$+^{30}$Mg$$&$0$&$	77.0272 	$&$	-19.0051 	$&$	-27.4707 	$&$	25.70 	$&$	25.94 	$&$	25.68 	$&$	26.59 	$&$	25.77 	$\\
	$^{238}$Pu$\to^{206}$Hg$+^{32}$Si$$&$0$&$	91.4554 	$&$	-20.3917 	$&$	-25.8916 	$&$	25.28 	$&$	25.75 	$&$	25.09 	$&$	29.55 	$&$	25.24 	$\\
	$^{242}$Cm$\to^{208}$Pb$+^{34}$Si$$&$0$&$	96.8219 	$&$	-20.7044 	$&$	-23.2721 	$&$	23.15 	$&$	23.46 	$&$	21.96 	$&$	28.74 	$&$	23.13 	$\\
		
\end{longtable*}
\endgroup

\begingroup
\renewcommand*{\arraystretch}{1.3}
\setlength{\tabcolsep}{1mm}
\setlength\LTleft{0pt}
\setlength\LTright{0pt}
\setlength{\LTcapwidth}{7in}
\begin{longtable*}
	{@{\extracolsep{\fill}} cccccccccc}
	\caption{Predicted half-lives for possible cluster radioactive nuclei.}
	\label{Tab3} \\
	\hline 
	\hline 
	{Cluster decay}&$\ell$&$Q_{c}$&$S_{c}$&$P$&
	$\rm{log}_{10}{\emph{T}}_{1/2}^{\rm{\,exp}}$&$\rm{log}_{10}{\emph{T}}_{1/2}^{\rm{\,cal}}$&$\rm{log}_{10}{\emph{T}}_{1/2}^{\rm{\,UDL}}$&$\rm{log}_{10}{\emph{T}}_{1/2}^{\rm{\,San}}$&$\rm{log}_{10}{\emph{T}}_{1/2}^{\rm{\,Ni}}$
	\\
	\hline
	\endfirsthead
	\multicolumn{10}{c}%
	{{\tablename\ \thetable{} continued from previous page}} \\
	\hline
	\hline 
	{Cluster decay}&$\ell$&$Q_{c}$&$S_{c}$&$P$&
	$\rm{log}_{10}{\emph{T}}_{1/2}^{\rm{\,exp}}$&$\rm{log}_{10}{\emph{T}}_{1/2}^{\rm{\,cal}}$&$\rm{log}_{10}{\emph{T}}_{1/2}^{\rm{\,UDL}}$&$\rm{log}_{10}{\emph{T}}_{1/2} ^{\rm{\,San}}$&$\rm{log}_{10}{\emph{T}}_{1/2}^{\rm{\,Ni}}$	
	\\
	\hline
	\endhead
	\hline  \\
	\endfoot
	\hline \hline
	\endlastfoot	
$^{219}$Rn$\to^{205}$Hg$+^{14}$C$$&$	3	$&$	28.2045 	$&$	-9.3106 	$&$	-31.1945 	$&$	-	$&$	19.79 	$&$	18.82 	$&$	15.80 	$&$	20.05 	$\\
$^{220}$Rn$\to^{206}$Hg$+^{14}$C$$&$	0	$&$	28.6452 	$&$	-9.2286 	$&$	-30.1550 	$&$	-	$&$	18.67 	$&$	17.68 	$&$	15.32 	$&$	17.82 	$\\
$^{216}$Ra$\to^{202}$Pb$+^{14}$C$$&$	0	$&$	26.3232 	$&$	-9.7120 	$&$	-37.0122 	$&$	-	$&$	26.01 	$&$	26.44 	$&$	19.74 	$&$	25.40 	$\\
$^{217}$Ra$\to^{203}$Pb$+^{14}$C$$&$	3	$&$	27.7672 	$&$	-9.5651 	$&$	-33.9485 	$&$	-	$&$	22.80 	$&$	22.46 	$&$	17.93 	$&$	23.14 	$\\
$^{218}$Ra$\to^{204}$Pb$+^{14}$C$$&$	0	$&$	28.8470 	$&$	-9.4112 	$&$	-31.5791 	$&$	-	$&$	20.28 	$&$	19.66 	$&$	16.67 	$&$	19.42 	$\\
$^{219}$Ra$\to^{205}$Pb$+^{14}$C$$&$	1	$&$	30.2554 	$&$	-9.2436 	$&$	-28.9125 	$&$	-	$&$	17.44 	$&$	16.24 	$&$	15.12 	$&$	17.67 	$\\
$^{220}$Ra$\to^{206}$Pb$+^{14}$C$$&$	0	$&$	31.1487 	$&$	-9.1266 	$&$	-27.2797 	$&$	-	$&$	15.70 	$&$	14.18 	$&$	14.21 	$&$	14.61 	$\\
$^{225}$Ra$\to^{211}$Pb$+^{14}$C$$&$	4	$&$	29.5772 	$&$	-9.3203 	$&$	-30.1227 	$&$	-	$&$	18.74 	$&$	17.57 	$&$	16.00 	$&$	19.13 	$\\
$^{227}$Ra$\to^{213}$Pb$+^{14}$C$$&$	4	$&$	27.4727 	$&$	-9.5527 	$&$	-34.2551 	$&$	-	$&$	23.11 	$&$	22.80 	$&$	18.52 	$&$	23.90 	$\\
$^{228}$Ra$\to^{214}$Pb$+^{14}$C$$&$	0	$&$	26.2145 	$&$	-9.6532 	$&$	-36.7678 	$&$	-	$&$	25.72 	$&$	26.24 	$&$	20.16 	$&$	25.76 	$\\
$^{226}$Th$\to^{212}$Po$+^{14}$C$$&$	0	$&$	30.6625 	$&$	-9.3402 	$&$	-29.5282 	$&$	>16.76	$&$	18.17 	$&$	17.28 	$&$	16.42 	$&$	17.59 	$\\
$^{221}$Fr$\to^{206}$Hg$+^{15}$N$$&$	3	$&$	34.2477 	$&$	-10.3874 	$&$	-31.9446 	$&$	-	$&$	21.64 	$&$	21.27 	$&$	19.35 	$&$	21.90 	$\\
$^{223}$Ac$\to^{208}$Pb$+^{15}$N$$&$	3	$&$	39.6028 	$&$	-9.9588 	$&$	-25.2032 	$&$	>14.76	$&$	14.47 	$&$	12.69 	$&$	16.14 	$&$	14.25 	$\\
$^{223}$Ra$\to^{205}$Hg$+^{18}$O$$&$	1	$&$	40.4495 	$&$	-12.4298 	$&$	-33.7972 	$&$	-	$&$	25.58 	$&$	26.13 	$&$	21.90 	$&$	26.13 	$\\
$^{225}$Ra$\to^{205}$Hg$+^{20}$O$$&$	1	$&$	40.6339 	$&$	-13.3417 	$&$	-34.5083 	$&$	-	$&$	27.22 	$&$	27.93 	$&$	21.07 	$&$	28.03 	$\\
$^{226}$Ra$\to^{206}$Hg$+^{20}$O$$&$	0	$&$	40.9665 	$&$	-13.2818 	$&$	-33.8685 	$&$	-	$&$	26.52 	$&$	27.13 	$&$	20.81 	$&$	26.11 	$\\
$^{227}$Ac$\to^{207}$Tl$+^{20}$O$$&$	1	$&$	43.2390 	$&$	-13.0992 	$&$	-31.1446 	$&$	-	$&$	23.62 	$&$	23.63 	$&$	19.79 	$&$	24.26 	$\\
$^{226}$Th$\to^{208}$Pb$+^{18}$O$$&$	0	$&$	45.8800 	$&$	-11.9610 	$&$	-27.3505 	$&$	-	$&$	18.67 	$&$	17.86 	$&$	19.24 	$&$	17.56 	$\\
$^{227}$Th$\to^{209}$Pb$+^{18}$O$$&$	4	$&$	44.3529 	$&$	-12.1921 	$&$	-29.6290 	$&$	>15.30	$&$	21.18 	$&$	20.70 	$&$	20.47 	$&$	21.43 	$\\
$^{229}$Th$\to^{209}$Pb$+^{20}$O$$&$	2	$&$	43.5582 	$&$	-13.1959 	$&$	-31.5903 	$&$	-	$&$	24.17 	$&$	24.31 	$&$	20.47 	$&$	24.96 	$\\
$^{227}$Pa$\to^{209}$Bi$+^{18}$O$$&$	2	$&$	46.0248 	$&$	-12.0873 	$&$	-28.0729 	$&$	-	$&$	19.52 	$&$	18.88 	$&$	20.01 	$&$	19.75 	$\\
$^{229}$Ac$\to^{206}$Hg$+^{23}$F$$&$	2	$&$	48.5111 	$&$	-15.1129 	$&$	-33.3699 	$&$	-	$&$	27.89 	$&$	28.58 	$&$	22.30 	$&$	28.83 	$\\
$^{228}$Th$\to^{206}$Hg$+^{22}$Ne$$&$	0	$&$	55.9268 	$&$	-15.1472 	$&$	-31.9105 	$&$	-	$&$	26.45 	$&$	27.14 	$&$	25.24 	$&$	25.76 	$\\
$^{229}$Th$\to^{205}$Hg$+^{24}$Ne$$&$	3	$&$	58.0103 	$&$	-15.7425 	$&$	-29.8573 	$&$	-	$&$	25.01 	$&$	25.00 	$&$	23.16 	$&$	25.42 	$\\
$^{231}$Th$\to^{207}$Hg$+^{24}$Ne$$&$	2	$&$	56.4396 	$&$	-15.9642 	$&$	-31.8145 	$&$	-	$&$	27.19 	$&$	27.77 	$&$	24.29 	$&$	28.04 	$\\
$^{231}$Th$\to^{206}$Hg$+^{25}$Ne$$&$	2	$&$	56.9830 	$&$	-16.3195 	$&$	-31.3968 	$&$	-	$&$	27.14 	$&$	27.56 	$&$	23.57 	$&$	27.99 	$\\
$^{232}$Th$\to^{208}$Hg$+^{24}$Ne$$&$	0	$&$	54.8535 	$&$	-16.1971 	$&$	-33.9632 	$&$	>29.20	$&$	29.58 	$&$	30.76 	$&$	25.44 	$&$	29.53 	$\\
$^{232}$Th$\to^{206}$Hg$+^{26}$Ne$$&$	0	$&$	56.0969 	$&$	-16.8859 	$&$	-32.8647 	$&$	>29.20	$&$	29.18 	$&$	30.01 	$&$	23.86 	$&$	29.12 	$\\
$^{229}$Pa$\to^{207}$Tl$+^{22}$Ne$$&$	2	$&$	59.1442 	$&$	-14.8614 	$&$	-28.8286 	$&$	-	$&$	23.09 	$&$	22.99 	$&$	24.07 	$&$	23.34 	$\\
$^{230}$U$\to^{208}$Pb$+^{22}$Ne$$&$	0	$&$	61.5800 	$&$	-14.6702 	$&$	-26.8616 	$&$	>18.20	$&$	20.93 	$&$	20.42 	$&$	23.50 	$&$	19.81 	$\\
$^{226}$U$\to^{202}$Pb$+^{24}$Ne$$&$	0	$&$	59.4134 	$&$	-15.9530 	$&$	-30.3525 	$&$	-	$&$	25.71 	$&$	26.01 	$&$	24.03 	$&$	24.77 	$\\
$^{227}$U$\to^{203}$Pb$+^{24}$Ne$$&$	1	$&$	59.9744 	$&$	-15.8542 	$&$	-29.5899 	$&$	-	$&$	24.85 	$&$	24.94 	$&$	23.70 	$&$	25.15 	$\\
$^{228}$U$\to^{204}$Pb$+^{24}$Ne$$&$	0	$&$	60.4732 	$&$	-15.7588 	$&$	-28.8952 	$&$	-	$&$	24.06 	$&$	24.00 	$&$	23.42 	$&$	23.12 	$\\
$^{229}$U$\to^{205}$Pb$+^{24}$Ne$$&$	1	$&$	61.1246 	$&$	-15.6438 	$&$	-28.0475 	$&$	-	$&$	23.10 	$&$	22.80 	$&$	23.05 	$&$	23.38 	$\\
$^{230}$U$\to^{206}$Pb$+^{24}$Ne$$&$	0	$&$	61.5439 	$&$	-15.5599 	$&$	-27.4727 	$&$	>18.20	$&$	22.45 	$&$	22.02 	$&$	22.82 	$&$	21.49 	$\\
$^{231}$U$\to^{207}$Pb$+^{24}$Ne$$&$	3	$&$	62.4014 	$&$	-15.4240 	$&$	-26.4674 	$&$	-	$&$	21.31 	$&$	20.50 	$&$	22.34 	$&$	21.47 	$\\
$^{235}$U$\to^{211}$Pb$+^{24}$Ne$$&$	1	$&$	57.5552 	$&$	-16.1488 	$&$	-32.3277 	$&$	>27.65	$&$	27.90 	$&$	28.80 	$&$	25.59 	$&$	29.19 	$\\
$^{236}$U$\to^{212}$Pb$+^{24}$Ne$$&$	0	$&$	56.1369 	$&$	-16.3590 	$&$	-34.2490 	$&$	-	$&$	30.04 	$&$	31.45 	$&$	26.62 	$&$	30.37 	$\\
$^{237}$U$\to^{213}$Pb$+^{24}$Ne$$&$	4	$&$	54.7375 	$&$	-16.5943 	$&$	-36.3721 	$&$	-	$&$	32.40 	$&$	34.17 	$&$	27.66 	$&$	34.14 	$\\
$^{238}$U$\to^{214}$Pb$+^{24}$Ne$$&$	0	$&$	53.6341 	$&$	-16.7265 	$&$	-37.8870 	$&$	-	$&$	34.05 	$&$	36.38 	$&$	28.52 	$&$	34.92 	$\\
$^{235}$U$\to^{210}$Pb$+^{25}$Ne$$&$	3	$&$	57.8750 	$&$	-16.5626 	$&$	-32.2852 	$&$	>27.65	$&$	28.28 	$&$	29.04 	$&$	25.03 	$&$	29.54 	$\\
$^{236}$U$\to^{212}$Pb$+^{24}$Ne$$&$	0	$&$	56.1369 	$&$	-16.3590 	$&$	-34.2490 	$&$	>26.27	$&$	30.04 	$&$	31.45 	$&$	26.62 	$&$	30.37 	$\\
$^{236}$U$\to^{210}$Pb$+^{26}$Ne$$&$	0	$&$	56.8838 	$&$	-17.1445 	$&$	-33.8876 	$&$	>26.27	$&$	30.47 	$&$	31.72 	$&$	25.41 	$&$	30.89 	$\\
$^{231}$Np$\to^{209}$Bi$+^{22}$Ne$$&$	3	$&$	62.0984 	$&$	-14.7961 	$&$	-27.2942 	$&$	-	$&$	21.49 	$&$	21.05 	$&$	24.14 	$&$	21.67 	$\\
$^{233}$Np$\to^{209}$Bi$+^{24}$Ne$$&$	3	$&$	62.3553 	$&$	-15.6205 	$&$	-27.4506 	$&$	-	$&$	22.49 	$&$	22.04 	$&$	23.36 	$&$	22.95 	$\\
$^{232}$U$\to^{204}$Hg$+^{28}$Mg$$&$	0	$&$	74.5456 	$&$	-18.1642 	$&$	-27.8405 	$&$	>22.26	$&$	25.44 	$&$	25.23 	$&$	26.35 	$&$	24.60 	$\\
$^{233}$U$\to^{205}$Hg$+^{28}$Mg$$&$	3	$&$	74.4532 	$&$	-18.1826 	$&$	-27.9368 	$&$	>27.59	$&$	25.56 	$&$	25.30 	$&$	26.44 	$&$	26.01 	$\\
$^{235}$U$\to^{207}$Hg$+^{28}$Mg$$&$	1	$&$	72.6519 	$&$	-18.4594 	$&$	-29.8019 	$&$	>28.45	$&$	27.71 	$&$	28.08 	$&$	27.52 	$&$	28.66 	$\\
$^{235}$U$\to^{206}$Hg$+^{29}$Mg$$&$	3	$&$	72.7034 	$&$	-18.8906 	$&$	-30.0074 	$&$	>28.45	$&$	28.35 	$&$	28.65 	$&$	27.15 	$&$	29.33 	$\\
$^{236}$U$\to^{208}$Hg$+^{28}$Mg$$&$	0	$&$	70.9607 	$&$	-18.7383 	$&$	-31.7358 	$&$	>26.27	$&$	29.92 	$&$	30.86 	$&$	28.53 	$&$	29.98 	$\\
$^{236}$U$\to^{206}$Hg$+^{30}$Mg$$&$	0	$&$	72.4982 	$&$	-19.3164 	$&$	-30.3402 	$&$	>26.27	$&$	29.11 	$&$	29.55 	$&$	26.97 	$&$	29.12 	$\\
$^{238}$U$\to^{208}$Hg$+^{30}$Mg$$&$	0	$&$	69.6853 	$&$	-19.8081 	$&$	-33.7060 	$&$	-	$&$	32.97 	$&$	34.37 	$&$	28.69 	$&$	33.61 	$\\
$^{235}$Np$\to^{207}$Tl$+^{28}$Mg$$&$	2	$&$	77.3271 	$&$	-17.9031 	$&$	-25.8705 	$&$	-	$&$	23.22 	$&$	22.47 	$&$	26.00 	$&$	23.61 	$\\
$^{237}$Np$\to^{207}$Tl$+^{30}$Mg$$&$	2	$&$	75.0172 	$&$	-19.1218 	$&$	-28.6050 	$&$	>27.57	$&$	27.19 	$&$	27.15 	$&$	26.63 	$&$	28.29 	$\\
$^{237}$Pu$\to^{209}$Pb$+^{28}$Mg$$&$	1	$&$	77.9604 	$&$	-18.0289 	$&$	-26.2041 	$&$	-	$&$	23.68 	$&$	23.14 	$&$	26.74 	$&$	24.34 	$\\
$^{237}$Pu$\to^{208}$Pb$+^{29}$Mg$$&$	3	$&$	77.6868 	$&$	-18.5058 	$&$	-26.7075 	$&$	-	$&$	24.67 	$&$	24.14 	$&$	26.54 	$&$	25.40 	$\\
$^{239}$Pu$\to^{209}$Pb$+^{30}$Mg$$&$	4	$&$	75.3184 	$&$	-19.3390 	$&$	-29.4080 	$&$	-	$&$	28.21 	$&$	28.39 	$&$	27.56 	$&$	29.54 	$\\
$^{237}$Am$\to^{209}$Bi$+^{28}$Mg$$&$	2	$&$	80.0868 	$&$	-17.9256 	$&$	-25.1679 	$&$	-	$&$	22.55 	$&$	21.70 	$&$	26.63 	$&$	23.03 	$\\
$^{237}$Pu$\to^{205}$Hg$+^{32}$Si$$&$	4	$&$	91.7259 	$&$	-20.3826 	$&$	-25.7756 	$&$	-	$&$	25.62 	$&$	24.78 	$&$	29.38 	$&$	26.13 	$\\
$^{239}$Pu$\to^{205}$Hg$+^{34}$Si$$&$	1	$&$	91.1365 	$&$	-21.2212 	$&$	-26.3681 	$&$	-	$&$	27.06 	$&$	26.42 	$&$	29.07 	$&$	28.14 	$\\
$^{239}$Am$\to^{207}$Tl$+^{32}$Si$$&$	3	$&$	94.7753 	$&$	-20.0817 	$&$	-23.9974 	$&$	-	$&$	23.55 	$&$	22.26 	$&$	29.14 	$&$	24.04 	$\\
$^{241}$Am$\to^{207}$Tl$+^{34}$Si$$&$	3	$&$	94.2333 	$&$	-20.9170 	$&$	-24.5466 	$&$	>24.41	$&$	24.94 	$&$	23.73 	$&$	28.80 	$&$	25.90 	$\\
$^{240}$Cm$\to^{208}$Pb$+^{32}$Si$$&$	0	$&$	97.8283 	$&$	-19.7873 	$&$	-22.3677 	$&$	-	$&$	21.63 	$&$	19.93 	$&$	28.88 	$&$	20.76 	$\\
$^{241}$Cm$\to^{209}$Pb$+^{32}$Si$$&$	4	$&$	95.6720 	$&$	-20.2239 	$&$	-24.2690 	$&$	-	$&$	23.97 	$&$	22.80 	$&$	29.87 	$&$	24.68 	$\\
$^{243}$Cm$\to^{209}$Pb$+^{34}$Si$$&$	2	$&$	95.0662 	$&$	-21.0605 	$&$	-24.8302 	$&$	-	$&$	25.38 	$&$	24.37 	$&$	29.56 	$&$	26.64 	$\\
$^{244}$Cm$\to^{210}$Pb$+^{34}$Si$$&$	0	$&$	93.4499 	$&$	-21.3677 	$&$	-26.2778 	$&$	-	$&$	27.13 	$&$	26.64 	$&$	30.35 	$&$	27.52 	$\\
	
\end{longtable*}
\endgroup

\section{Summary}
\label{sec:Summary}
In summary, based on Wentzel-Kramers-Brillouin theory, considering the screened electrostatic effect of Coulomb potential, we systematically calculate the cluster radioactivity half-lives of 22 nuclei by using a phenomenological model. The calculated results obtained by using our model are satisfactory agreement with the experimental data. Moreover, we extend this model to predict the cluster radioactivity half-lives of 66 possible cluster radioactive candidates. It is found that the predictions have a good consistency with several contrasting models and/or formulas. Finally, the low values of the cluster radioactivity half-lives at daughter neutron number $(N_{d}=126)$ confirms the important role of neutron magicity. It also shows that
the neutron shell closure is more influential than the proton shell closure in the process of cluster radioactivity. This work may provide useful help for the future experimental research on cluster radioactivity.

\begin{acknowledgments}

This work is supported in part by the National Natural Science Foundation of China (Grants No. 12175100 and No.11975132), the Construct Program of the Key Discipline in Hunan Province, the
Research Foundation of Education Bureau of Hunan Province, China (Grant No. 21B0402 and No. 18A237), the Natural Science Foundation of Hunan Province, China (Grants No. 2018JJ2321), the
Innovation Group of Nuclear and Particle Physics in USC, the Shandong Province Natural Science Foundation, China (Grant No. ZR2022JQ04), the Hunan Provincial Innovation Foundation For Postgraduate (Grant No. CX20220993) and the Opening Project of Cooperative Innovation Center for Nuclear Fuel Cycle Technology and Equipment, University of South China (Grant No. 2019KFZ10).
\end{acknowledgments}


%

\end{document}